\input{aipcheck}

\documentclass[
    ,final            
  ]{aipproc}

\layoutstyle{8x11single}

\begin{document}

\title{Nuclear energy density functionals: what we
can learn about/from their global performance?}



\author{A. V. Afanasjev}{
  address={Department of Physics and Astronomy, Mississippi
  State University, MS 39762, USA}
}

\author{S. E. Agbemava}{
  address={Department of Physics and Astronomy, Mississippi
  State University, MS 39762, USA}
}

\author{D. Ray}{
  address={Department of Physics and Astronomy, Mississippi
  State University, MS 39762, USA}
}

\author{P. Ring}{
  address={Fakult\"at f\"ur Physik, Technische Universit\"at M\"unchen,
 D-85748 Garching, Germany}
}

\begin{abstract}
A short review of recent results on the global performance of covariant 
energy density functionals is presented. It is focused on an analysis 
of the accuracy of the description of physical observables of ground 
and excited states as well as to related theoretical uncertainties. In 
addition, a global analysis of pairing properties is presented and 
the impact of pairing on the position of two-neutron drip line is 
discussed.
\end{abstract}

\maketitle


\section{1. Introduction}

To fulfill specific requirements of nuclear structure and nuclear
astrophysics (such as an high demand for the reliability of 
theoretical predictions), the {\it microscopic}  and {\it universal} 
aspects of nuclear theory should be contemplated. A {\it microscopic} 
description by a physically sound model ensures a reliable 
extrapolation away from the experimentally known region. A {\it universal} 
description of all nuclear properties within one unique framework for 
all nuclei involved ensures a coherent prediction of all unknown data. 
The quest towards such a theoretical tool will most certainly be in 
the focus of fundamental nuclear physics research in the foreseeable 
future. At present, the tool of choice for the description of medium 
heavy and heavy nuclei is density functional theory (DFT)
\cite{Koh.99,LNP.641,Drut2010_PPNP64-120}:  for the majority of such 
nuclei there is simply no other microscopic alternative. Among these 
nuclear DFT's, covariant density functional theory (CDFT) is one of 
most attractive since {\it covariant energy density functionals}
(CEDF)  exploit basic properties of QCD at low energies, such as
symmetries and the separation of scales \cite{LNP.641,VALR.05}. They
also provide a consistent treatment of the spin degrees of freedom
\cite{VALR.05}, spin-orbit interactions \cite{BRRMG.99,LA.11} and
time-odd mean fields \cite{KR.93,AA.10}.

It is important to understand the accuracy of the description of
the properties of the ground and excited states in nuclei, theoretical
uncertainties in the description of related physical observables and
the predictive power of the models on a global scale. This is especially
crucial for nuclear astrophysics, where we are facing the problem of
an extrapolation to nuclei with large isospin. Many of such nuclei
will not be studied experimentally even with the next generation of
facilities, or forever. The advent of high-performance computers has
recently allowed to address these questions. As illustrated in Refs.\
\cite{AARR.13,AARR.14}, such a global analysis is feasible for ground
state observables. On the
other hand, for spectroscopic observables such a systematic analysis
of the accuracy of the description of experimental data and related
theoretical uncertainties has been performed only for actinides
\cite{AS.11,AO.13,A.14} because of the complexity of their calculations.

The paper is organized as follows. Some representative examples
of the studies of the accuracy of the description of physical
observables of ground and excited states and related
theoretical uncertainties are presented in Sects. 2 and 3,
respectively. Considering that detailed results of these studies
have already been published we focus on new results in Sects. 4
and 5. Sect. 4 shows the results of the studies of the global
numerical accuracies due to the truncation of the basis in the
calculations of binding energies. Sect. 5 presents the results of
on-going study of the pairing properties and their impact on the
position of two-neutron drip line. Finally, Sect.\ 6 summarizes
the results of our work.

\section{2. Global performance and the estimate of theoretical uncertainties}

\begin{figure*}[ht]
\includegraphics[width=16.0cm,angle=0]{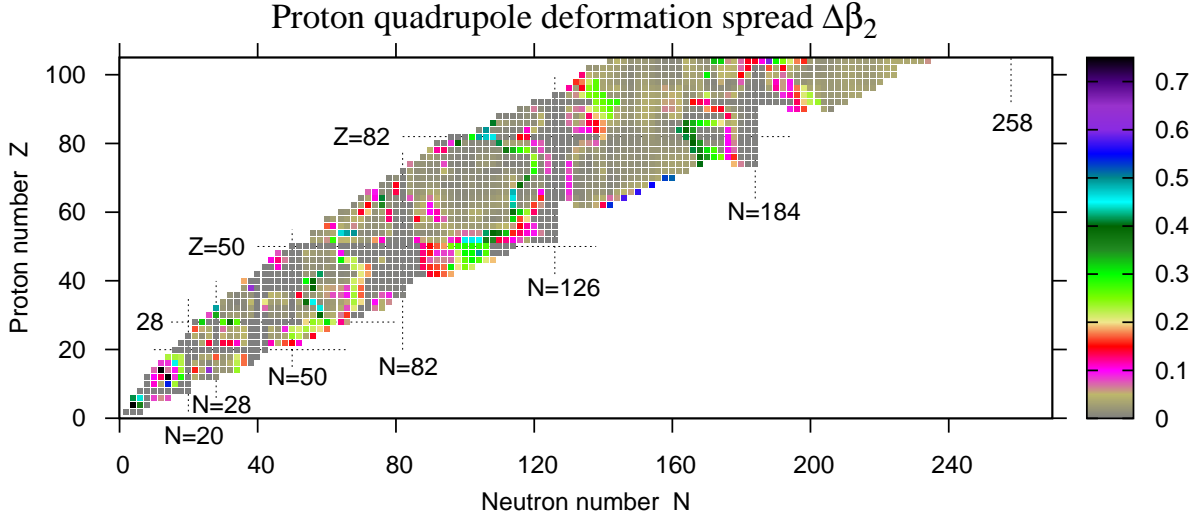}
\caption{Proton quadrupole deformation spreads
$\Delta \beta_2(Z,N)$ as a function of proton and neutron number.
$\Delta \beta_2(Z,N)=|\beta_2^{\rm max}(Z,N)-\beta_2^{\rm min}(Z,N)|$,
where $\beta_2^{\rm max}(Z,N)$ and $\beta_2^{\rm min}(Z,N)$ are the
largest and smallest proton quadrupole deformations obtained
with four employed CEDF's (NL3*, DD-ME2, DD-ME$\delta$ and DD-PC1)
for $(Z,N)$ nuclei at the ground state.}
\label{Edif-charge-deform}
\end{figure*}

The first ever extensive study of global performance of CEDF's
has been performed in Refs.\ \cite{AARR.13,AARR.14}. There were two
main goals behind this study. First was the assessment of global
performance of the state-of-the art CEDF's with respect of the
description of ground state properties of even-even nuclei. The
following physical observables were analysed: binding energies,
charge radii, neutron skin thicknesses, deformations, the positions
of the two-proton and two-neutron drip lines. The analysis of these
results will allow in future to define better strategies for new
fits of CEDF's. The second goal was to estimate theoretical
uncertainties in the description of various physical observables
on a global scale and especially in the regions of unknown nuclei.

It was suggested in Refs.\ \cite{RN.10,DNR.14} to use
the methods of information theory and to define the uncertainties
in the energy density functional (EDF) parameters. These
uncertainties come from the selection
of the form of EDF as well as from the fitting protocol details,
such as the selection of the nuclei under investigation, the
physical observables, or the corresponding weights. Some of them
are called {\it statistical errors} and can be calculated from a
statistical analysis during the fit, others are systematic errors,
such as for instance the form of the EDF under investigation. On the
basis of these statistical errors and under certain assumptions on
the independence of the form of many EDF's one hopes to be able to
deduce in this way
{\it  theoretical error bars}
for the prediction
of physical observables~\cite{RN.10,DNR.14}.
The calculation
of properties of transitional and deformed nuclei requires a considerable
amount of computer time and therefore it is
difficult to perform the analysis of statistical errors on a global
scale since the properties of all nuclei have
to be calculated repeatedly for different variations of
original CEDF. Thus, such statistical analysis has been performed
mostly for spherical nuclei \cite{RN.10,KENBGO.13}
or selected isotopic chains of deformed nuclei \cite{Eet.12}.

As a consequence, we concentrate mostly on the
uncertainties related to the present choice of EDF's
which can be relatively easily deduced globally. We
therefore define {\it theoretical systematic uncertainties}
for a given physical observable via the spread of theoretical
predictions within the four CEDF's (namely, NL3* \cite{NL3*},
DD-ME$\delta$ \cite{DD-MEdelta}, DD-ME2 \cite{DD-ME2} and
DD-PC1 \cite{DD-PC1})
\begin{equation}
\Delta O(Z,N)=|O_{\rm max}(Z,N)-O_{\rm min}(Z,N)|
\label{eq:TSUC}
\end{equation}
where $O_{\rm max}(Z,N)$ and $O_{\rm min}(Z,N)$ are the largest
and smallest values of the physical observable $O(Z,N)$ obtained
with the four employed CEDF's for the $(Z,N)$ nucleus. Note that 
these {\it theoretical uncertainties} are only spreads of physical 
observables due to a very small number of functionals and, thus, 
they are only a crude approximation to the {\it systematic 
theoretical errors} discussed in Ref.~\cite{DNR.14}.

\begin{figure*}[ht]
\includegraphics[width=14.0cm]{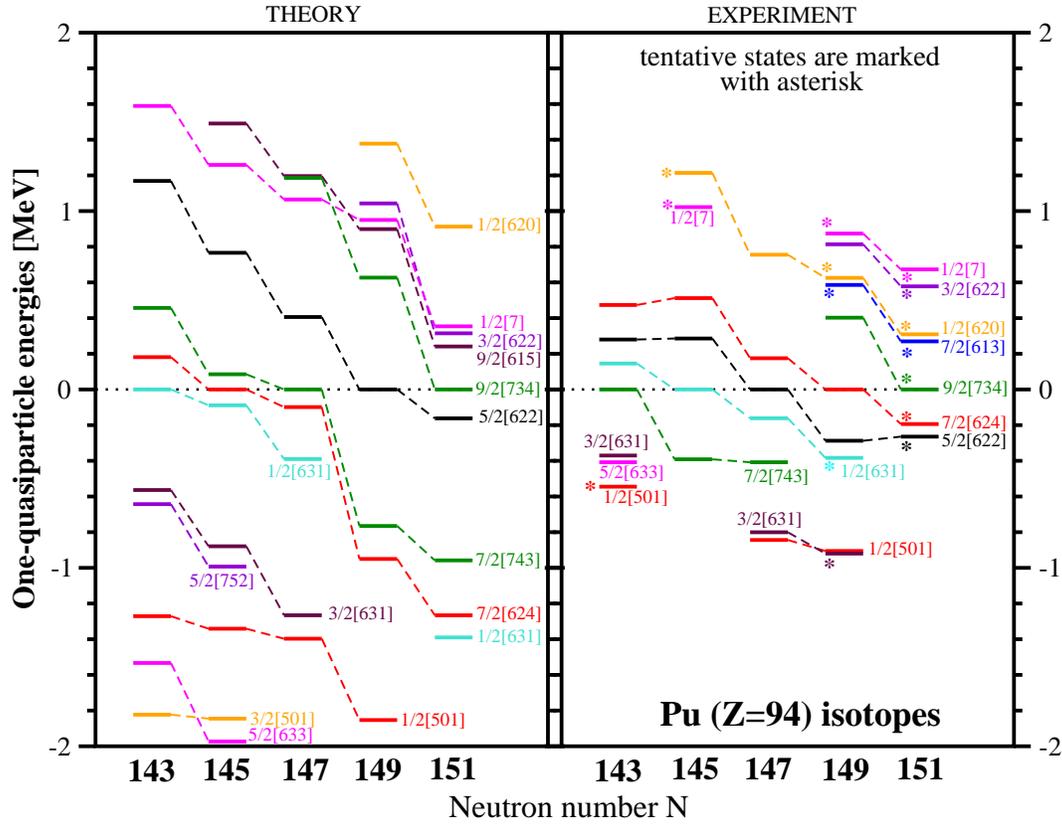}
\vspace{-0.5cm}
\caption{The evolution of one-quasineutron energies as a function of neutron
number for the Pu isotopes. Hole states are plotted below the ground state
(zero energy), and particle states are plotted above. Experimental data
(one-quasineutron band-head energies) are taken from Ref.\ \cite{Eval-data}.
The states are labelled by the Nilsson labels $\Omega [Nn_z\Lambda]$ or by
only principal quantum number $N$ and $\Omega$ in the case of strongly mixed
wave function.}
\label{Eqp-evol-neu}
\end{figure*}

 Although such an analysis has its own merits, at present, it does 
not allow to estimate real theoretical error bars in the description 
of physical observables. This is because they originate not only
from the uncertainties in model parameters, but also from the
definition and the limitations of the model itself.
As in the case of present Skyrme functionals, the different CEDF's
are far from forming an independent statistical ensemble.
Their number is very small and they are all based on a very
similar form. For example, no tensor terms are present in the
relativistic functionals under investigation and simple power laws
are used  for the density dependence  in the Skyrme DFT. The parameters 
of these functionals are fitted according to similar protocols 
including similar types of physical observables such as binding 
energies and radii. In addition, there exist principal theoretical 
deficiencies for all presently used nuclear density functionals. 
Although there exist 
formal mathematical existence theorems for exact density functionals 
in an external field~\cite{Hohenberg1964_PR136-B864}, a similar theorem 
has not been proven for self-bound systems such as 
nuclei~\cite{Engel2007_PRC75-014306,Giraud2008_PRC77-014311,Nakatsukasa2012_PTEP01A207}.
In addition, there is the problem of shape coexistence in transitional nuclei, 
where the superposition of nuclei with different shapes has to be described 
by a linear combination of Slater determinants, i.e. by the methods going 
beyond mean field, and, therefore, beyond the concept of conventional density 
functional theory~\cite{NVR.11}. The later uncertainties are 
very difficult to estimate. As a consequence, any analysis of theoretical 
uncertainties (especially, for extrapolations to neutron-rich nuclei) contains 
a degree of arbitrariness related to the choice of the model and fitting 
protocol in detail and to the concept of density functional theory in nuclei 
in general.

Fig.\ \ref{Edif-charge-deform} shows an example of such analysis
of theoretical uncertainties; in this case ground state proton
quadrupole deformations are considered. Theoretical uncertainties
for this physical observable are either non-existent or very small
for spherical or nearly spherical nuclei as well as for well-deformed
nuclei in the rare-earth and in the actinide regions. The largest
spreads for predicting the equilibrium quadrupole  deformations
exist at the boundaries between regions of different deformations.
They are extremely high in the regions of the prolate-oblate shape
coexistence, indicating that the ground state in a given nucleus
can be prolate (oblate) in one CEDF and oblate (prolate) in
another CEDF. These uncertainties are due to the deficiencies of
the current generations of the DFT models with respect of the
description of single-particle energies \cite{AARR.14}.

\begin{figure*}[ht]
\centering
\includegraphics[width=16cm,angle=0]{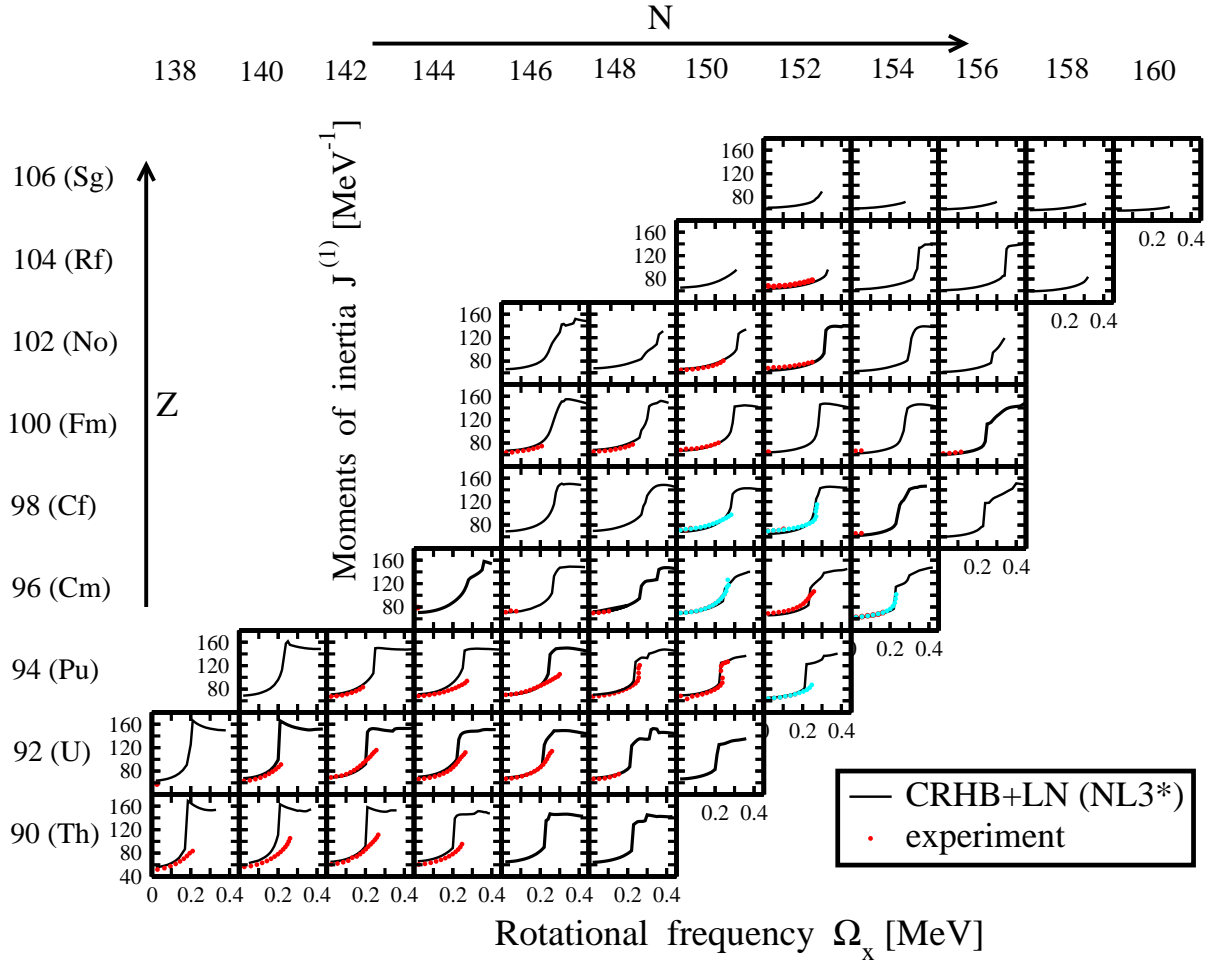}
\vspace{-0.8cm}
\caption{Experimental and calculated moments
of inertia $J^{(1)}$ as a function of rotational frequency $\Omega_x$.
The calculations are performed with the NL3* CEDF \cite{NL3*}. Calculated
results and experimental data are shown by black lines and red dots,
respectively. Cyan dots are used for new data (Ref.\ \cite{Hota.thesis})
which became available after publication of Ref.\ \cite{AO.13}. From
Ref.\ \cite{A.14}.}
\label{sys-J1-NL3s}
\end{figure*}

\section{3. Systematic studies in specific regions of nuclear chart}

As illustrated in Fig.\ \ref{Edif-charge-deform}, the theoretical
uncertainties in the prediction of ground state proton quadrupole
deformations are rather small for the regions of well deformed
nuclei such as rare-earth and actinides. Moreover, the experimental
data on $\beta_2$ in these regions are well (typically within the
experimental uncertainties) described by CDFT (see Sect.\ IX in
Ref.\ \cite{AARR.14}). As a result, such experimental data cannot
be used to differentiate between the functionals. It turns out that
the biggest difference between the CEDF's exists for the single-particle
spectra (see, for example, Ref.\ \cite{A250}).

Fig.\ \ref{Eqp-evol-neu} shows an example of the comparison between
calculated and experimental one-quasineutron  spectra in Pu isotopes
\cite{AS.11}. This reference represent most comprehensive study
(among any DFT) of the single-particle excitations in deformed
systems. A number of features are clearly seen.  First, for a given 1-qp state
the discrepancy between theory and experiment depends on neutron number.
Second, for a given 1-qp state the slope of the energy
curve as a function of neutron number is more pronounced in the calculations than
in experiment. These two features are interconnected and they emerge from the fact
that theoretical energy scale is more stretched out than the experimental one
due to the low effective mass. The change of the Fermi
energy with neutron number leads  to changes of the energy differences between
ground and excited states and these differences are affected by the effective mass in the
calculations. Third, the relative energies of different experimental
1-qp states are not always reproduced in the calculations. This feature originates
from the fact that the energies of spherical subshells, from which the deformed states
emerge, can deviate from experiment \cite{A250}). These three
features are seen in all isotone and isotope chains.

It is also interesting to see how other spectroscopic observables are
described in experiment and how they depend on the choice of the CEDF.
Fig.\ \ref{sys-J1-NL3s} shows the results of the first ever (in
any DFT framework) systematic investigation of rotational properties
of even-even nuclei at normal deformation \cite{AO.13}. The calculations
are performed within the cranked relativistic Hartree-Bogoliubov
(CRHB) approach with approximate particle number projection by
means of the Lipkin-Nogami method (further CRHB+LN) \cite{CRHB}.
One can see that the moments of inertia below band crossings are
reproduced well. The upbendings observed in a number of rotational
bands of the $A\geq 242$ nuclei are also reasonably well described
in the model calculations. Moreover, the CRHB+LN approach has a good
predictive power as illustrated by the fact that new data (cyan
dots) are very close to model predictions. However, the calculations
also predict similar upbendings in lighter nuclei, but they have not
been seen in experiment. The analysis suggests that the stabilization
of octupole deformation at high spin, not included in the present
CRHB+LN calculations, could be responsible for this discrepancy
between theory and experiment \cite{AO.13}. With few exceptions, the
rotational properties of one-quasiparticle configurations, which yield
important information on their underlying structure and, thus, provide
an extra tool for configuration assignment, are also well described
in the CRHB+LN calculations (see Ref.\ \cite{AO.13} for details).

\section{4. The estimate of numerical uncertainties in the calculations
         of binding energies}

\begin{figure}
  \includegraphics[width=8cm,angle=-90]{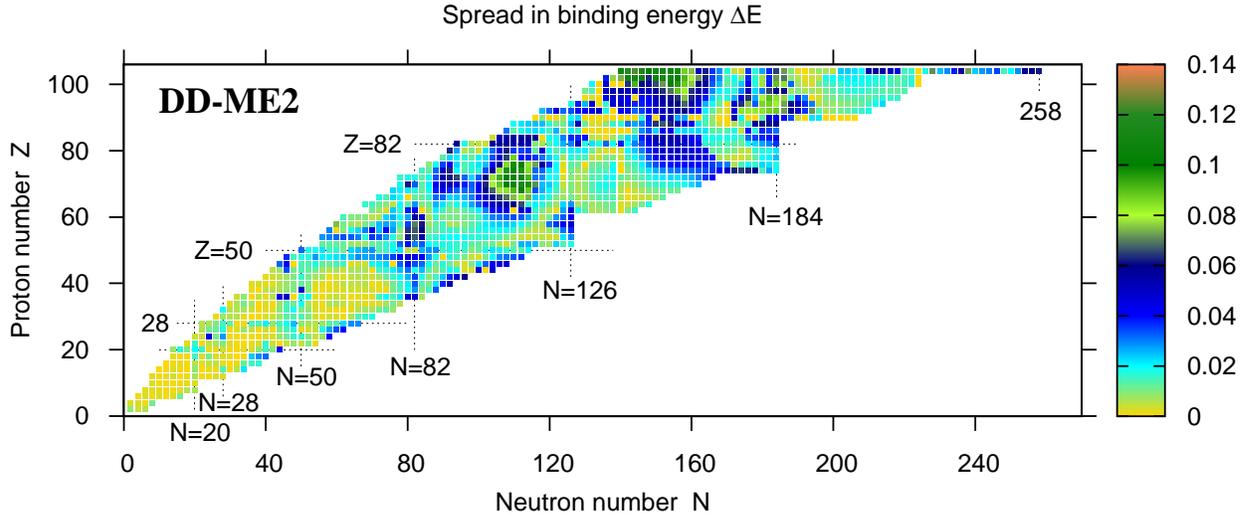}
  \caption{The spread of calculated binding energies when
           different deformations of the basis are used
           in the calculations. See text for details.}
\label{spreads}
\end{figure}

In the RHB calculations of Ref.\ \cite{AARR.14}, the truncation of
the basis is performed in such a way that all states belonging to
the major shells up to $N_F = 20$ fermionic shells for the Dirac
spinors and up to $N_B = 20$ bosonic shells for the meson fields
are taken into account. Note that a similar truncation of the basis
is usually used in the fitting protocols of the state-of-the-art CEDF's.
For each nucleus the potential energy curve is calculated in a
large deformation range from $\beta_2=-0.4$ up to $\beta_2=1.0$
by means of the constraint on the quadrupole moment $q_{20}$.
In constrained calculations, the deformation of the basis is
selected in such a way that it corresponds to the desired
deformation of the converged solution. The lowest in energy minimum
is defined from  the potential energy curve. Then, unconstrained
calculations are performed in this minimum and the correct ground
state configuration and its energy are determined. This procedure
is especially important for the cases of shape coexistence.

It is important to estimate how the selection of the basis
affects the calculated binding energies in the unconstrained
calculations. In order to address this question the calculations
for all even-even nuclei under study have been performed with the
DD-ME2 CEDF employing eight values of the deformation of the basis
starting from $\beta_2=-0.3$ and increasing up to $\beta_2=0.4$ in
step of 0.1. Then in each nucleus the spread of calculated binding
energies has been calculated as a difference of the minimal and maximal
binding energies obtained in the ground state minimum. These spreads
are shown in Fig.\ \ref{spreads}. One can see that they are below 20
keV in the absolute majority of light nuclei. These spreads are higher
in medium and heavy mass nuclei. However, even there they are below
100 keV.

  Unconstrained calculations for a given deformation of basis do
not always converge to the ground state minimum. This typically
takes place in transitional nuclei with soft potential energy surfaces
and nuclei with prolate-oblate shape coexistence. In these nuclei,
the calculated binding energies of only a few deformations of the basis
enter in the definition of the spreads. As a result, these spreads are
smaller than in well-deformed nuclei. On the other hand, in well-deformed
nuclei almost all employed deformations of the basis lead to the ground
state minimum in the calculations. That is a reason why among $Z\geq 50$
nuclei the spreads of binding energies are most pronounced in the
well-deformed nuclei of the rare-earth region and the actinides
(Fig.\ \ref{spreads}).

This analysis clearly shows that the employed truncation of basis
provides accurate results; extremely small numerical errors exist
in light nuclei but they increase somewhat with the increase of the 
mass of the nucleus. Note that the same approach has been used in
unconstrained calculations of Ref.\ \cite{AARR.14}; the binding energy
of the ground state configuration has been defined as a minimum energy
of the unconstrained calculations obtained for eight deformations of
the basis. It was also observed that not always the lowest in energy
solution is generated by the deformation of the basis which is close
to the deformation of the solution.

\section{5. Pairing properties}

\begin{figure}
\includegraphics[width=7cm,angle=-90]{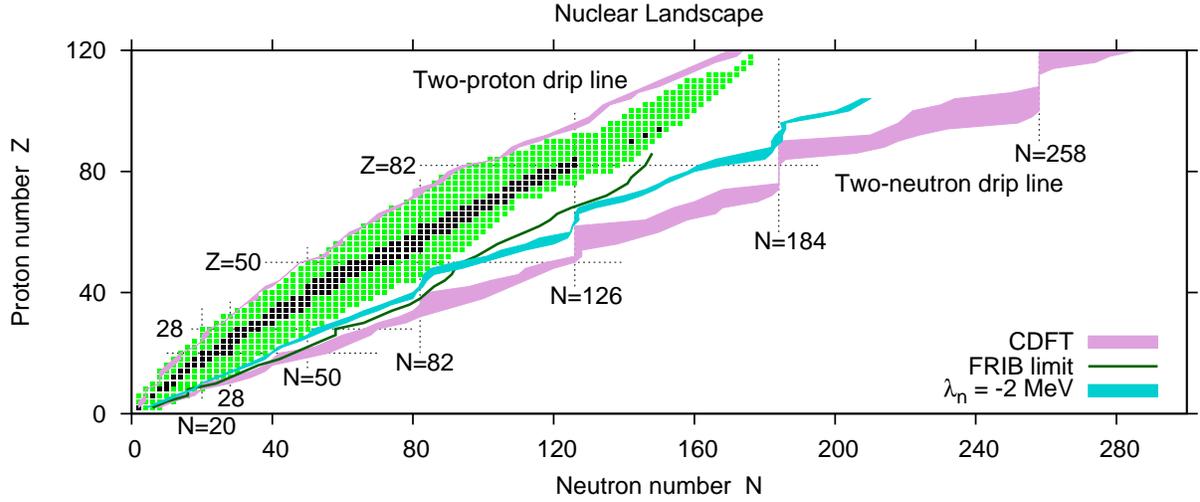}
\caption{Nuclear landscape as provided by state-of-the-art
CDFT calculations. The uncertainties in the definition of
two-proton and two-neutron drip lines are shown by violet
shaded areas. They are defined by the extremes of the
predictions of the corresponding drip lines obtained with
different functionals. The uncertainties (the range of
nuclei) in the definition of the neutron chemical potential
$\lambda_n=-2.0$ MeV are shown by blue shaded area.
Experimentally known stable and radioactive (including proton 
emitters) nuclei  are shown by black and green squares, respectively. 
Green solid line shows the limits of the nuclear chart (defined as 
fission yield greater than $10^{-6}$) which may be achieved with 
dedicated
existence measurements at FRIB \protect\cite{S-priv.14}.
Based on Fig. 4 of Ref.\ \protect\cite{AARR.13}.}
\label{landscape}
\end{figure}

\begin{figure}
  \includegraphics[width=18cm,angle=0]{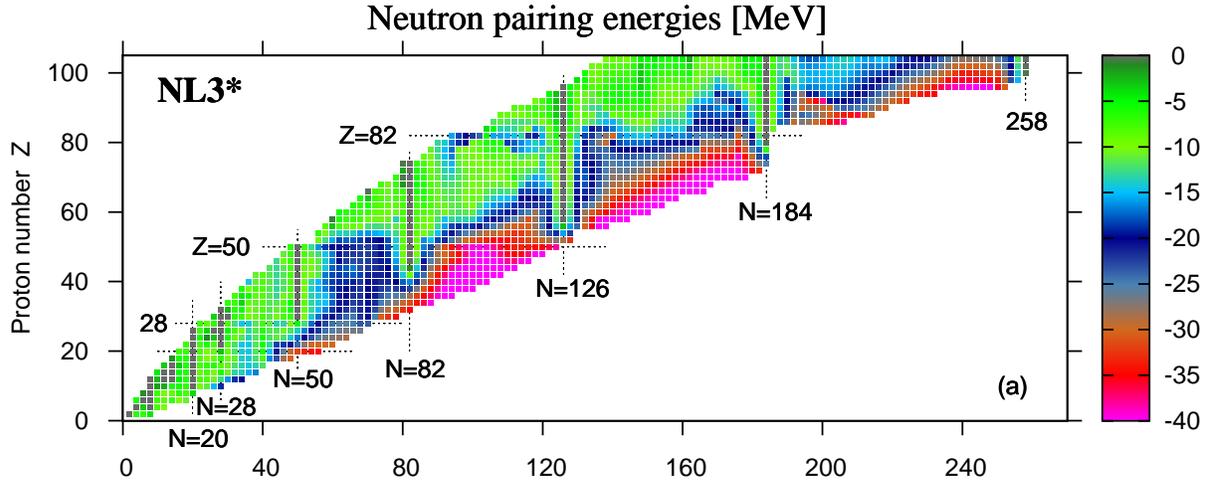} 
  \caption{Neutron pairing energies $E_{pairing}$ obtained
           in the RHB calculations with the NL3* CEDF.}
\label{neu_pair_nl3s}
\end{figure}

\begin{figure}
  \includegraphics[width=18cm,angle=0]{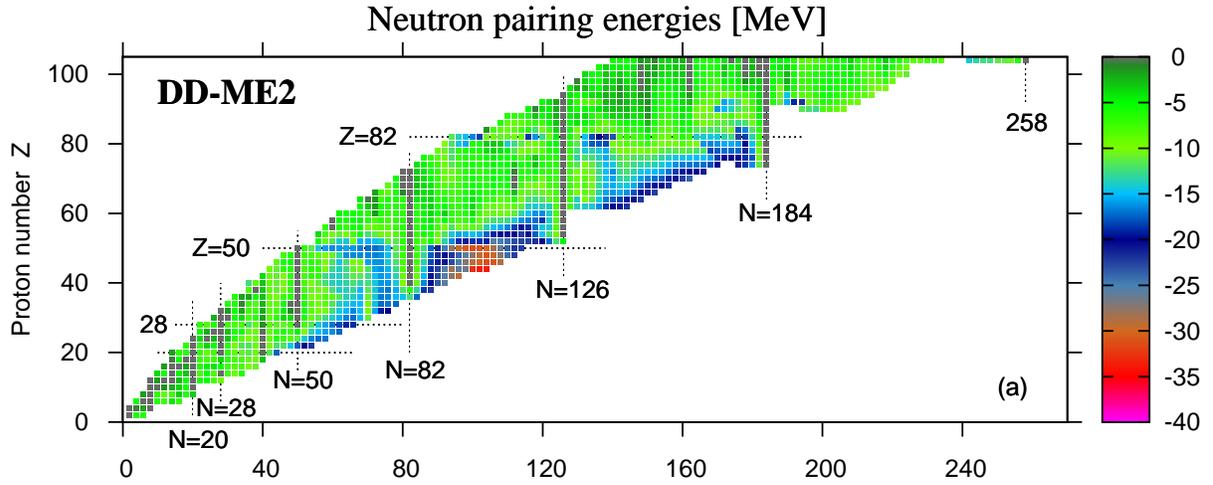}
  \caption{The same as Fig.\ \ref{neu_pair_nl3s} but for the DD-ME2 CEDF.}
\label{neu_pair_ddme2}
\end{figure}

Other interesting questions are the evolution of pairing properties
with isospin, their dependence on the CEDF's, the uncertainties in their
definition and the impact of these factors on the physical observables
of interest. There are two measures of pairing correlations, namely,
the pairing gap $\Delta$ and the pairing energies $E_{pairing}$. In addition,
the energies of proton or neutron chemical potentials and their evolution
with particle number are important for the definition of the positions of
the proton and neutron drip lines and the regions of the nuclear chart where
the coupling with the continuum may become important. Fig.\ \ref{landscape}
shows the nuclear landscape as obtained with four state-of-the-art CEDF's
(NL3*, DD-ME2, DD-PC1 and DD-ME$\delta$) and related theoretical uncertainties
in the definition of the two-proton and two-neutron drip lines \cite{AARR.13}.
In addition, this figure compares the uncertainties (the range of nuclei)
in the definition of the neutron chemical potential $\lambda_n=-2.0$ MeV (blue
shaded area) with a possible extension (green solid line) of the nuclear
landscape by means of the facility for the rare isotope beams (FRIB). These 2
MeV represents roughly the typical size of the energy window above and below
the chemical potential in which the pairing interaction affects measurably the
occupation of the single-particle orbitals. If the neutron chemical potential
of a given nucleus is closer than 2 MeV to the continuum, the pairing interaction
can scatter pairs to the continuum. Thus, Fig.\
\ref{landscape} suggests that in future FRIB experiments the region of
nuclei with measurable coupling with the continuum can be reached only in
the $Z\leq 50$ nuclei. For higher $Z$ nuclei, future experimental data on
neutron-rich nuclei can be safely treated without accounting of the coupling
with the continuum.

\begin{figure}
  \includegraphics[width=14cm,angle=0]{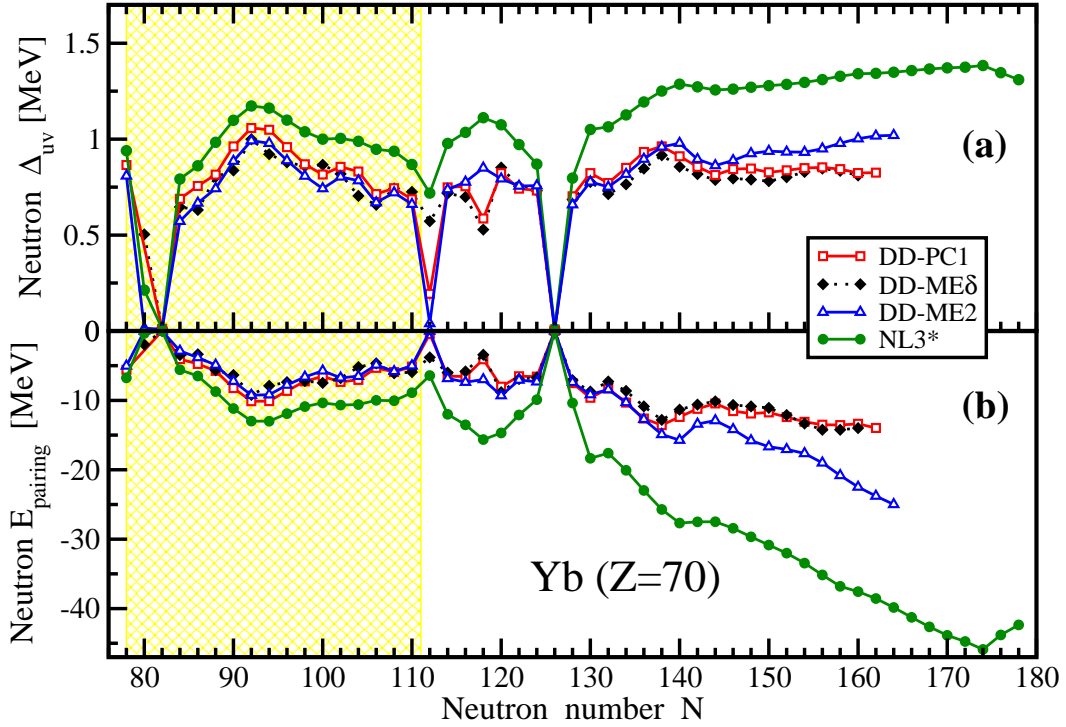}
  \caption{Neutron pairing gaps $\Delta_{uv}$ and
           pairing energies $E_{pairing}$ of the Yb nuclei
           located between the two-proton and two-neutron
           drip-lines obtained in the axial RHB calculations
          with the indicated CEDF's. The shaded yellow
          area indicates experimentally known nuclei.}
\label{Yb-result}
\end{figure}

 At present, several definitions of the pairing gap exist \cite{AARR.14}.
However, the analysis presented in Sect.\ IV of Ref.\ \cite{AARR.14}
clearly indicates that the $\Delta_{\rm uv}$ values for the pairing gap
defined in even-even nuclei provide the best agreement with the pairing
indicators calculated from odd-even mass staggerings. The pairing gap
\begin{equation}
\Delta_{\rm uv}=\frac{\sum_k u_kv_k\Delta_k}{\sum_k u_kv_k}
\end{equation}
averages over $u_kv_k$, a quantity which is concentrated around
the Fermi surface.
In Hartree-(Fock)-Bogoliubov calculations the size of the
pairing correlations is usually measured in terms of the
pairing energy defined as
\begin{eqnarray}
E_{pairing}~=~-\frac{1}{2}\mbox{Tr} (\Delta\kappa).
\label{Epair}
\end{eqnarray}
This is not an experimentally accessible quantity, but it is a measure
for the size of the pairing correlations in theoretical calculations.

Figs.\ \ref{neu_pair_nl3s} and \ref{neu_pair_ddme2} compare
neutron pairing energies $E_{pairing}$ obtained with the NL3*
and DD-ME2 CEDF's. In the region of known nuclei these energies
are only slightly smaller (but, in general, quite comparable)
in DD-ME2 as compared with NL3*. However, on approaching the
two-neutron drip line, substantial differences develop between
the pairing energies in the RHB calculations with these two
CEDF's. The absolute values of neutron pairing energies obtained
in RHB calculations with NL3* are by factor of 3-4 higher near the
two-neutron drip line than those in known the nuclei (Fig.\
\ref{neu_pair_nl3s}). This difference reduces to a factor 2 for
the DD-ME2 CEDF (Fig.\ \ref{neu_pair_ddme2}). The results obtained with DD-ME$\delta$
and DD-PC1 are similar to the ones obtained with DD-ME2.

  Unfortunately, at present similar figures cannot be generated for
the $\Delta_{uv}$ pairing gaps because of an improper recording of
this quantity during the production calculations for Ref.\
\cite{AARR.14}. Thus, we perform the comparison of the evolution
of neutron pairing gaps $\Delta_{uv}$ and pairing energies
$E_{pairing}$ as a function of neutron number only for the Yb
$Z=70$ isotope chain (see Fig.\ \ref{Yb-result}). One can see
that in the RHB calculations with the DD-ME$\delta$, DD-ME2 and
DD-PC1 CEDF's the pairing gaps $\Delta_{uv}$ in neutron-rich
$N\geq 126$ nuclei have on average the same magnitude as pairing
gaps in known nuclei (Fig.\ \ref{Yb-result}a). However, the absolute
pairing energies are larger by a factor of about 2 in neutron-rich
nuclei as compared with the ones in known nuclei. Note that both
$\Delta_{uv}$ and $E_{pairing}$ are more or less constant in
neutron-rich nuclei in the RHB calculations with DD-PC1 and
DD-ME$\delta$. On the contrary, some increase of absolute
values of these quantities with increasing isospin is observed
in DD-ME2.

\begin{figure}
  \includegraphics[width=14cm,angle=0]{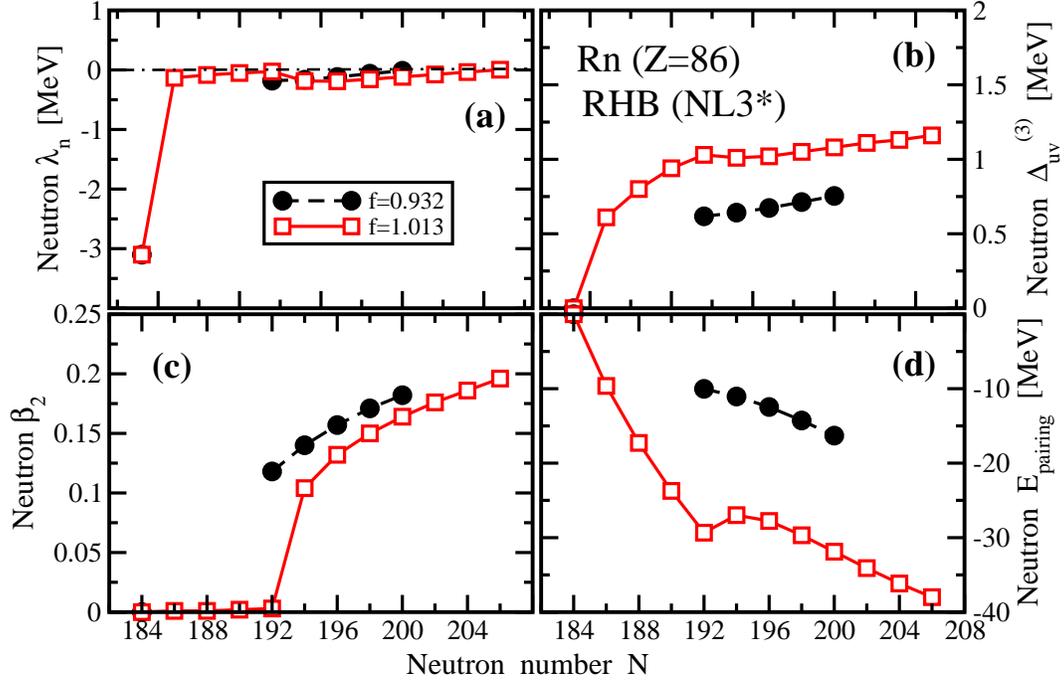}
  \caption{The evolution of neutron chemical potential
  $\lambda_n$ (panel (a)), neutron quadrupole deformation
  $\beta_2$ (panel (c)), neutron pairing gap $\Delta_{uv}$
  (panel (b)) and neutron pairing energy $E_{pairing}$
  (panel (d)) as a function of neutron number $N$ in the
  Rn isotopes with $N\geq 184$ obtained
  in the RHB (NL3*) calculations. Only the results for bound
  nuclei are shown. The results of the calculations for two
  values of scaling factor $f$ of separable pairing force of
  finite range are presented. Note that the value of
  $f=1.013$ has been used in the calculations of Ref.\
  \protect\cite{AARR.14}.
}
\label{drip-line}
\end{figure}

The situation is different in the NL3* CEDF. The pairing is
somewhat stronger in known nuclei as compared with DD CEDF's.
However, more pronounced differences are seen when the results
in neutron-rich nuclei are compared with the ones in known
nuclei. The pairing gaps $\Delta_{uv}$ are on average 25\%
larger in neutron-rich nuclei as compared with known ones and,
in addition, they gradually increase with neutron number. The
absolute values of the pairing energies rapidly increase with
neutron number in neutron-rich $N\geq 126$ nuclei; near two-proton
drip line these energies are larger by a factor of 4 than average
pairing energies in known nuclei.

Note that in Ref.\ \cite{AARR.14}, the selection of scaling
factors $f$ for separable pairing has been guided by the comparison
of experimental data with different calculations
based on the NL3* CEDF. The same scaling factors $f$ were
used also in the calculations with DD-PC1, DD-ME2 and DD-ME$\delta$.
The spread in the calculated values $\Delta_{uv}$ values in
known nuclei indicates that the scaling factors $f$ used in Ref.\
\cite{AARR.14} are reasonable within few \%
(see Sec.\ IV in Ref.\ \cite{AARR.14} and Fig.\ \ref{Yb-result}
in the present paper).
The weak dependence of the scaling factor $f$ on the CEDF has
already been seen in the studies of pairing and rotational
properties in the actinides ~\cite{A250,AO.13}. Considering
the global character of the study in Ref.\ \cite{AARR.14}, this
is a reasonable choice. Definitely there are nuclei in which
the choice of the scaling factors $f$ is not optimal.

Considering the differences in the predictions of the
pairing properties of the nuclei near the drip line, it is important
to understand how they affect physical observables of
interest such as the position of two-neutron drip line.
To address this question we analyze the chain of Rn
$(Z=86)$ isotopes, the two-neutron drip lines of which
are located at $N=206$ for NL3* and at $N=184$ for
DD-ME2, DD-ME$\delta$, and DD-PC1 in the calculations
of Ref.\ \cite{AARR.14}.

First, we perform the RHB(NL3*) calculations with a pairing
strength decreased by 8\% (scaling factor $f=0.932$) as
compared to the one used in Ref.\ \cite{AARR.14}. This will
roughly put the calculated pairing energies near the two-proton drip
line into the range close to the one obtained in the calculations
with DD-ME$\delta$, DD-ME2 and DD-PC1 CEDF's (compare Figs.\
\ref{neu_pair_ddme2} and \ref{drip-line}d). This decrease of
pairing strength has a significant effect on the nuclei near
the two-neutron drip line and the position of the two-neutron
drip line. Indeed, the Rn isotopes with $N=186, 188, 190, 202, 204$
and 206, which are bound for the original strength of pairing
(scaling factor $f=1.013$), become unbound for decreased pairing.
In addition, the deformations of the $N=192-200$ isotopes become
larger (Figs.\ \ref{drip-line}c); this reflects the well known
fact that pairing typically tries to make a nucleus less
deformed.

 Second, we try to bring the situation in the RHB calculations
with the DD CEDF's close to the one seen in NL3* by increasing
the pairing strength by 8\% in the RHB calculations with DD-ME2
and DD-PC1. However, this does not affect the position of
two-neutron drip lines for the Rn isotope chain in these CEDF's
because of the details of the underlying shell structure. These
two results clearly show that the actual position of the
two-neutron drip line sensitively depends on fine details of
the underlying shell structure and the strength of pairing.

\section{6. Conclusions}

During the last several years  considerable progress in an
understanding of the global performance of state-of-the-art
covariant energy density functionals and related theoretical
uncertainties has been achieved. Many physical observables
related to the ground state properties (binding energies,
charge radii, deformations, neutron skin thicknesses,
the positions of drip lines etc), the properties of excited
states (moments of inertia, energies of predominantly
single-particle states, fission barriers \cite{AAR.10} etc) 
have been studied either globally or at least systematically 
in a specific region of nuclear chart. Theoretical uncertainties
for many physical observables have been defined. The current 
generation of CEDF's has been fitted only to bulk and nuclear 
matter properties. The existing discrepancies between theory 
and experiment clearly indicate the need for the inclusion of 
single-particle information into the fitting protocol of next 
generation of CEDF's. It is doubtful that without inclusion of 
such information missing terms (and missing physics) of CEDF's 
can be identified.


\begin{theacknowledgments}

This work has been supported by the U.S. Department of Energy
under the grant DE-FG02-07ER41459 and by the DFG cluster of excellence
\textquotedblleft Origin and Structure of the Universe
\textquotedblright\ (www.universe-cluster.de). This work was also
partially supported  by an allocation of advanced computing resources
provided by the National Science Foundation. The computations were
partially performed on Kraken at the National Institute for
Computational Sciences (http://www.nics.tennessee.edu/).

\end{theacknowledgments}



\bibliographystyle{aipproc}   

\bibliography{references}

\IfFileExists{\jobname.bbl}{}
 {\typeout{}
  \typeout{******************************************}
  \typeout{** Please run "bibtex \jobname" to optain}
  \typeout{** the bibliography and then re-run LaTeX}
  \typeout{** twice to fix the references!}
  \typeout{******************************************}
  \typeout{}
 }

\end{document}